\documentclass[aps,pra,showpacs, longbibliography]{revtex4-2}

\usepackage{graphicx}
\usepackage{dcolumn}
\usepackage{bm}
\usepackage{color}
\usepackage{amsmath}
\usepackage{amssymb}
\usepackage{comment}
\usepackage{xcolor}

\begin{document}


\title{Controlling and measuring a superposition of position and momentum}

\author{Takafumi Ono$^{1,2,3}$}
 \email{ono.takafumi@kagawa-u.ac.jp}

\author{Nigam Samantarray$^{1,4}$}%
 \email{nigam.samantaray@strath.ac.uk }

\author{John G. Rarity$^{1}$}
 \email{John.Rarity@bristol.ac.uk}
 \affiliation{%
 $^{1}$Quantum Engineering Technology Labs, H. H. Wills Physics Laboratory 
 and Department of Electrical \& Electronic Engineering, University of Bristol,
 Bristol BS8 1FD, UK\\
}%

\affiliation{%
$^{2}$Program in Advanced Materials Science
Faculty of Engineering and Design,
Kagawa University,
2217-20 Hayashi-cho, Takamatsu, Kagawa
761-0396, Japan\\
} 

\affiliation{%
$^{3}$JST, PRESTO, 4-1-8 Honcho, Kawaguchi, Saitama 332-0012, Japan\\
} 

\affiliation{%
 $^4$Department of Physics, University of Strathclyde, John Anderson Building, 107 Rottenrow, Glasgow G4 0NG, UK
 }
\date{\today}
 
\begin{abstract}
The dynamics of a particle propagating in free space is described by its position and momentum, where quantum mechanics prohibits the simultaneous identification of two non-commutative physical quantities. 
Recently, a lower bound on the probability of finding a particle after propagating for a given time has been derived for well-defined initial constraints on position and momentum under the assumption that particles travel in straight lines.
Here, we investigate this lower limit experimentally with photons.
We prepared a superposition of position and momentum states by using slits, lenses and an interferometer, and observed a quantum interference between position and momentum.
The lower bound was then evaluated using the initial state and the result was 5.9\% below this classical bound.
\end{abstract}

\maketitle
While the dynamics of a particle in free space is characterized by its position and momentum, the uncertainty principle prohibits identifying non-commuting observables of position and momentum simultaneously in quantum mechanics\cite{Heisenberg,Wootters1979,Scully1991,Brukner1999,Jacques2007,Sen2014}. There have been reports on the direct measurement of the wave function \cite{Lundeen2011} and the identification of particle trajectories using weak measurements \cite{Kocsis2011,Zhou2017}, and particle trajectories based on the Feynman path integral form have been analyzed \cite{Sinha2010,Sawant2014,Magana-Loaiza2016}. However, the interpretation of these identified trajectories and their physical meaning are still controversial. In particular, these results give the impression that the classical laws of motion of free particles do not hold at all.

On the other hand, in recent years, paradoxical aspects of particle propagation in free space have been discussed \cite{Hofmann2017, Hofmann2021}.
According to Heisenberg's equation of motion, the time evolution of the position of a free particle still has the same form as Newton's first law, and so it is still natural to assume that the particle moves in a straight line in free space.
In this context, an inequality of particle propagation has been derived based on the assumption of the classical laws of motion that the particle moves along a straight line.
Specifically, probability distributions are assigned to the non-commutative positions and momenta in Heisenberg's equations of motion at a given initial time and a lower bound on the probability of finding a particle at some later time has been derived using the assumption that particles travel in straight lines. An inequality then arises when the initial state consists of a superposition of two states one with well defined initial position and one with a well defined range of momenta. When these two interfere this reduces the probability of finding the particle within the range predicted by straight line paths at intermediate times probability \cite{Hofmann2017}.

In this letter, we experimentally prepared the superposition of position and momentum, and then investigated how much the quantum coherent process between position and momentum can deviate from the statistical limit which is predicted by the experimentally obtained statistical information of initial position and the momentum.
We prepared the superposition state by using slits, lenses and interferometer and then measured the quantum state at intermediate time where two states are almost overlapped.
We have successfully observed a quantum interference between position and momentum where the obtained interference visibility was about 85 $\%$.
From the probabilities of finding the particle in initial position and momentum, we predict the minimal probabilities of finding the particle at intermediate time based on the assumption that the particle moves along straight lines. 
However, we confirmed that 5.9 $\%$ probability were not able to be explained by the statistical limit.

Let us first summarize the propagation inequality of a particle in free space discussed in ref.\cite{Hofmann2017}. In quantum mechanics, time evolution of position operator $\hat{x}(t)$ for a single particle is described by the Heisenberg's formulation. In terms of initial position $\hat{x}(0)$ and momentum $\hat{p}_x$, the time evolution of position operator in free space is given by
\begin{equation}
\label{Heisenberg}
    \hat{x}(t) = \hat{x}(0) + \frac{\hat{p}_x}{m}t.
\end{equation}
If we replace these operators with the concrete values of $x$ and $p_x$, this equation corresponds to Newton's first law, which states that a particle moves in a straight line in time.
However, due to the uncertainty relation in quantum mechanics, the concrete values of the non-commutative observables ($\hat{x}$ and $\hat{p}_x$) cannot be identified simultaneously.
In Ref\cite{Hofmann2017} a lower bound on the probability of finding a particle at an intermediate time was derived by assigning separate probability distribution to the non-commutative observables of position and momentum at initial time.
Specifically, if a particle is assumed to be moving in a straight line, then a particle existing in an interval of width $-L/2 < x <L/2$ with probability of $P(L)$ and also in an interval $-B/2<p_x<B/2$ at $t=0$ with probability of $P(B)$ must pass through the position interval $M=L+Bt/m$ with probability of $P(M, t)$. This constraint requires $P(M, t) \geq P(L~\rm{AND}~B)$.
The lowest possible value of $P(L~\rm{AND}~B)$ can be expressed using the experimentally observable probabilities of finding the particle in $L$ and $B$, resulting in the statistical limit of finding the particle at intermediate time as follows \cite{Hofmann2017},
\begin{equation}
\label{propagation_ineq}
    P(M,t) \geq P(L) + P(B) -1.
\end{equation}
Because $P(M,t)$, $P(L)$ and $P(B)$ are obtained from the actual statistics of position and momentum in a different time, this equation can thus be confirmed in an experiment.

Many quantum states violate the inequality in the equation, but in this work we have studied the originally considered states in ref\cite{Hofmann2017}, which is a superposition of a position state $| L \rangle$ localized in the spatial interval $L$ and a momentum state $| B \rangle$ localized in the momentum interval $B$ which is given by
\begin{equation}
\label{superposition}
| \psi(t=0) \rangle = \frac{1}{\sqrt{2( 1 + \langle L | B \rangle} )} \left( | L \rangle + | B \rangle \right).
\end{equation}
Note that the constructive interference between a position state $| L \rangle$ and a momentum state $| B \rangle$ enhances total probability $P(L) + P(B)$ at $t=0$ in eq.(\ref{propagation_ineq}).
With time, the shape of the position wave function $\langle x| \hat{U} | L \rangle$  broadens and approaches the sinc function, while the shape of the momentum wave function $\langle x| \hat{U} | B \rangle$ does not change with time.
The exact form of the wave functions for $t > mL^2/(2\pi \hbar)$ are given by
\begin{eqnarray}
\langle x | \hat{U} | L \rangle = & & \sqrt{\frac{mL}{2\pi \hbar t}} \left[ \frac{2\hbar t}{mLx} \sin \left(\frac{mL}{2\hbar t}x \right) \right] \nonumber\\
& & \times {\rm exp} \left( i \frac{m}{2\hbar t}x^2 - i \frac{\pi}{4} \right) \\
\langle x | \hat{U} | B \rangle = & & \sqrt{\frac{B}{2\pi \hbar}} \left[ \frac{2\hbar}{Bx} \sin \left(\frac{B}{2\hbar}x \right) \right].
\end{eqnarray}
A strong violation of the inequality can be observed at around $t_M = mL/B$ when the shapes of the wave function of position and momentum match perfectly, where strong quantum interference between position and momentum states reduces the probability of $P(M, t)$ in eq.(\ref{propagation_ineq}).

\begin{figure}[b]
 \centering
 \includegraphics[keepaspectratio,scale=0.5]{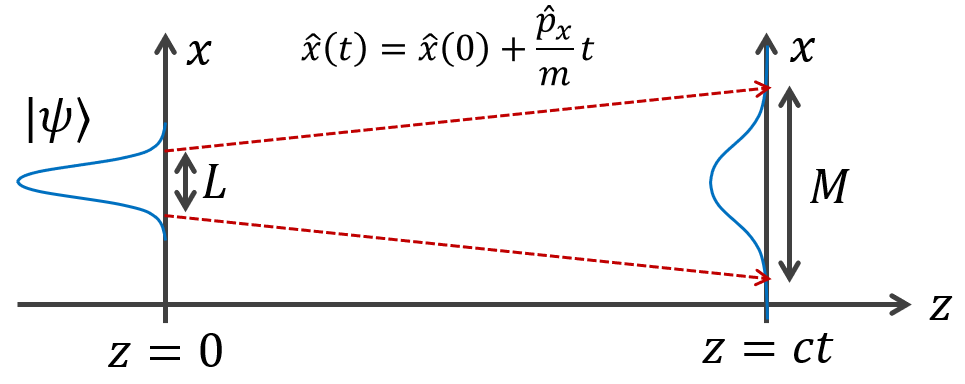}
 \caption{Conceptual illustration of the propagation inequality. Blue line shows a quantum state. Particle moves along $z$-axis. $\hat{p}_x$ corresponds to the transverse momentum of the photon. Red dotted line shows the boundary of the trajectory where photons in the initial position interval $L$ and momentum $B$ moves through the interval $M$.}
 \label{fig1}
\end{figure}
In this context, we have experimentally investigated the inequality of eq.(\ref{propagation_ineq}) with photons.
We utilized a transverse position $x$ and momentum $p_x$ of photons propagating at velocity $c$ along the $z$-axis,
where the position $z = ct$ is used to measure the time $t$ (Fig.1).
Using the paraxial approximation of $p_z \approx p = h/\lambda$, the transverse position can be identified as $\hat{x}(t) = \hat{x}(0) + (\hat{p}_x/p) \times ct$, so that the effective mass in eq.(\ref{Heisenberg}) can be translated into $m=p/c=h/(c\lambda)$ where $\lambda$ is the wavelength of photons.
In this configuration, the initial quantum state $|\psi \rangle$ at $t=0$ corresponds to the state at $z=0$, and the state at $t$ corresponds to the state at $z=ct$.

\begin{figure}[t]
 \centering
 \includegraphics[keepaspectratio,scale=0.4]{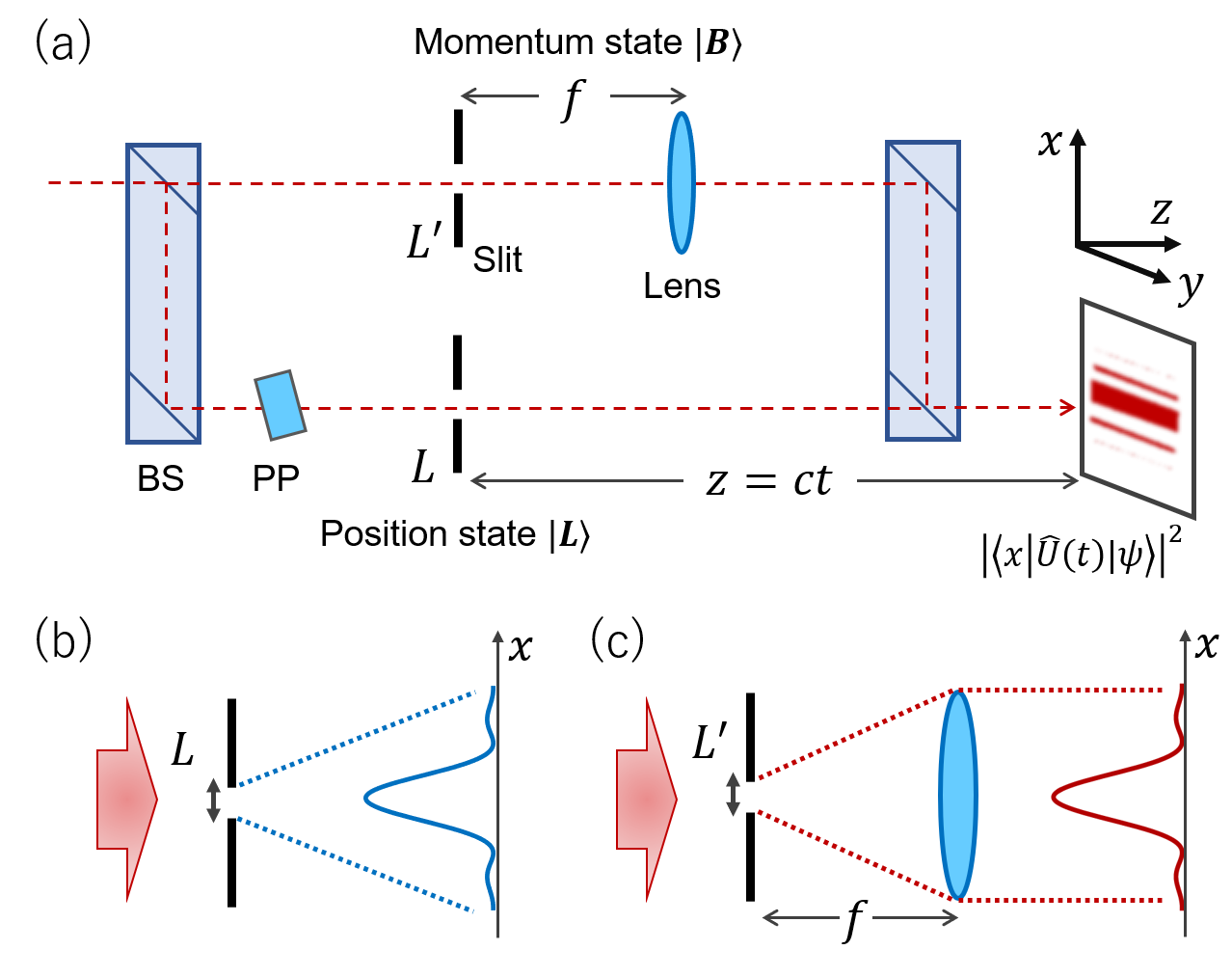}
 \caption{(a) Experimental setup for producing the superposition of position and momentum. BS: Beam splitter, PP: Phase plate. (b) The position state $| L \rangle$ was prepared by using slit $L$ in the bottom arm of the interferometer and (c) the momentum state $| B \rangle$ was prepared by using slit $L'$ and the lens $f$ in the upper arm of the interferometer. The lens $f'$ was used to create an image of the slit $L$ at the position of $2f'$ from the lens $f'$.}
 \label{fig1}
\end{figure}

Figure 2(a) shows the experimental setup for creating the superposition state of eq.(\ref{superposition}), which is based on Mach-Zehnder interferometer equipped with a slit and a lens.
The photon is input to the interferometer and divided by the first beam splitter, resulting in superposition of the photons in one path and in the other path.
To produce a position state $| L \rangle$, we put a slit of width $L$ in one arms of the interferometer (fig.2b), so that the state after the slit approximately corresponds to an image of the slit.
Likewise, to produce a momentum state, we used an effect on Fourier transform of the position by using a lens. 
We put a slit of width $L'$ in the other arm of the interferometer and the lens was placed on the position after the slit at the distance of focal length $f$ of the lens (fig.2c).
Under the paraxial approximation, the momentum information before the lens is mapped onto the position scaled by $h/(f \lambda)$ after the lens.
The corresponding momentum $B$ is therefore given by $B = h L'/(f\lambda$).
The state then corresponds to a superposition of position $|L\rangle$ and momentum $|B \rangle$.

\begin{figure}[t]
 \centering
 \includegraphics[keepaspectratio,scale=0.4]{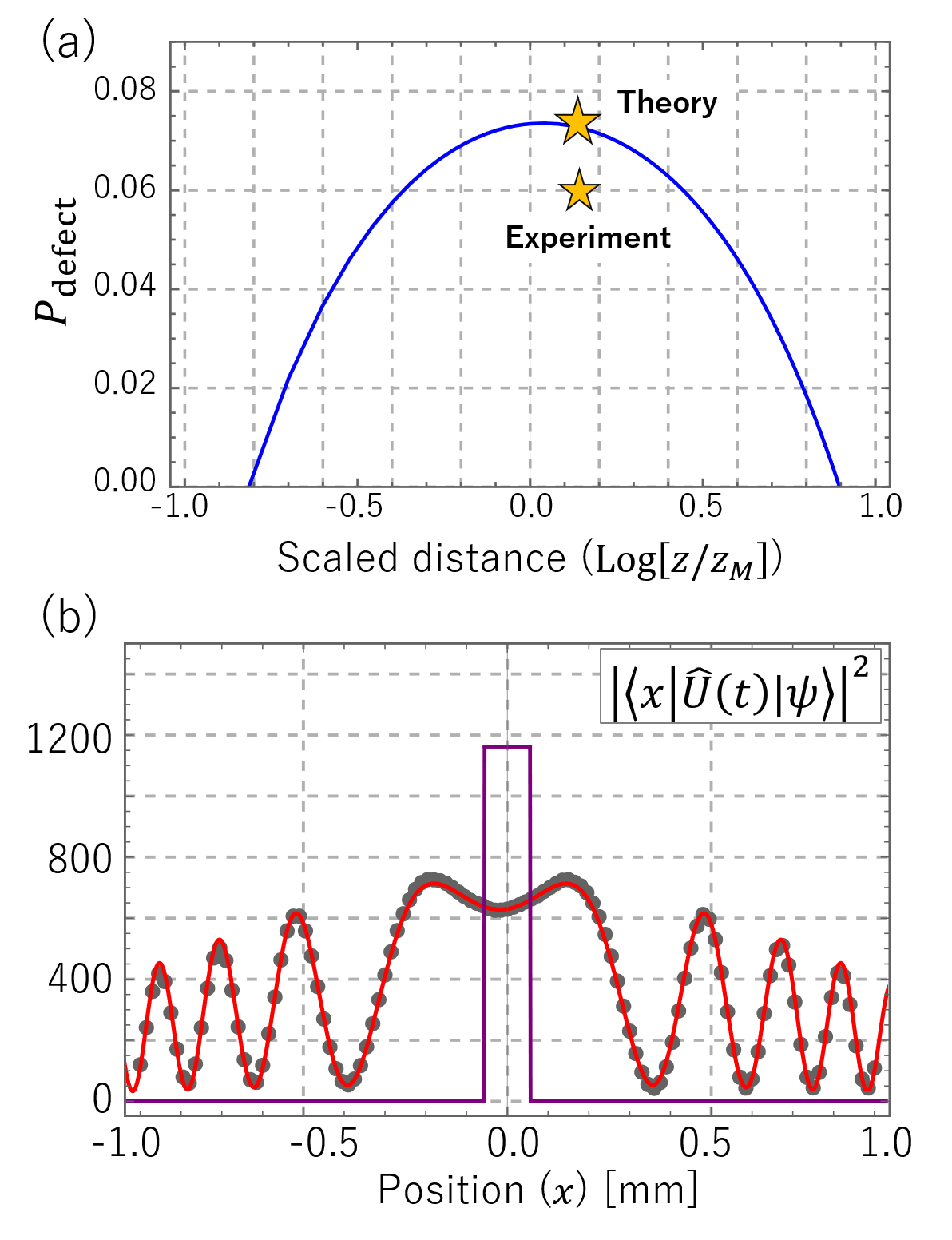}
 \caption{(a) Dependence of the defect probability on the normalized position of the photon scaled by $z_M$. The star-shaped points represent the conditions of the present experiment. The degradation of the experimental values with respect to the theoretical values is due to the visibility of the interference fringes.
 (b) Experimentally observed interference fringe between position and momentum at $z = 1.4 z_M$. Grey dot shows the experimental data. Red line shows the fitted curve. Purple line show the probability densities corresponding to the minimal value of $P(M)$ which is expected from eq.(\ref{propagation_ineq}). $ | \psi (t) \rangle = \hat{U}(t) | \psi \rangle$, where $\hat{U}(t)$ is unitary operator of time evolution in free space.
}
 \label{fig3}
\end{figure}

In order for the inequality in eq. (\ref{propagation_ineq}) to be strongly violated, $t$, $L$ and $B$ must be chosen carefully. 
To quantify the violation of the eq.(\ref{propagation_ineq}) at different time, we define the defect probability as
\begin{equation}
\label{defect}
    P_{\mathrm{defect}}(t) = P(L) + P(B) - 1 - P(M, t). 
\end{equation}
The condition for the inequality to be maximally violated is obtained by maximizing this probability.
As shown in \cite{Hofmann2017}, the maximum value of this probability is obtained when $LB=0.024 \times (2\pi \hbar)$ with $t_M=mL/B$ where the shapes of the wave functions of position and momentum match perfectly.
In the experiment, we chose the slit widths of position and momentum state for $L= 47 \mu \rm{m}$ and $L' = 37 \mu \rm{m}$ respectively, and we chose the lens of $f=10$ cm.
$LB$ was approximately calculated as $0.022 \times (2 \pi \hbar)$.
The position $z_M = c \times t_M$ where the shapes of the wave functions of position and momentum match perfectly is calculated as $10$ cm.

Figure 3(a) shows the dependence of the defect probability on the position scaled by $z_M$.
As shown in the figure, the maximum value of $P_{\rm defect}$ is obtained at around $z=z_M$.
More precisely, the maximum value of $P_{\rm defect}$ is obtained at a time $z=1.1 z_M$ which is slightly different from $z=z_M$ in our choice of $L$ and $B$.
Note that for our experimental parameters of slit widths and lens we have chosen, the optimal condition is different from the original ones in \cite{Hofmann2017}.
Note also that the violation of the inequality can be observed over a fairly wide time range from about $z\sim z_M$ to $z\sim 8z_M$, which correspond to the position from $z = 10$ cm to $z = 80$ cm.

In this proof of principle demonstration, we used a weak coherent laser light with wavelength of 800 nm. 
The laser was attenuated by using Neutral Density filter at a single photon level. 
The average photon number existing in the interferometer was estimated as 0.6 which ensure that the contribution of more than two photons from the laser was negligible. 
The interferometer was stable enough during the experiment due to the displaced common path design shown in figure 2(a).
The obtained visibility of our interferometer was about 85 \% as discussed later. 
We used a single photon sensitivity charge coupled device (CCD) to measure the interference fringe which ensures that the detected interference fringe is constructed from a pseudo-single photon source of weak coherent light.

\begin{figure*}[t]
 \centering
 \includegraphics[keepaspectratio,scale=0.4]{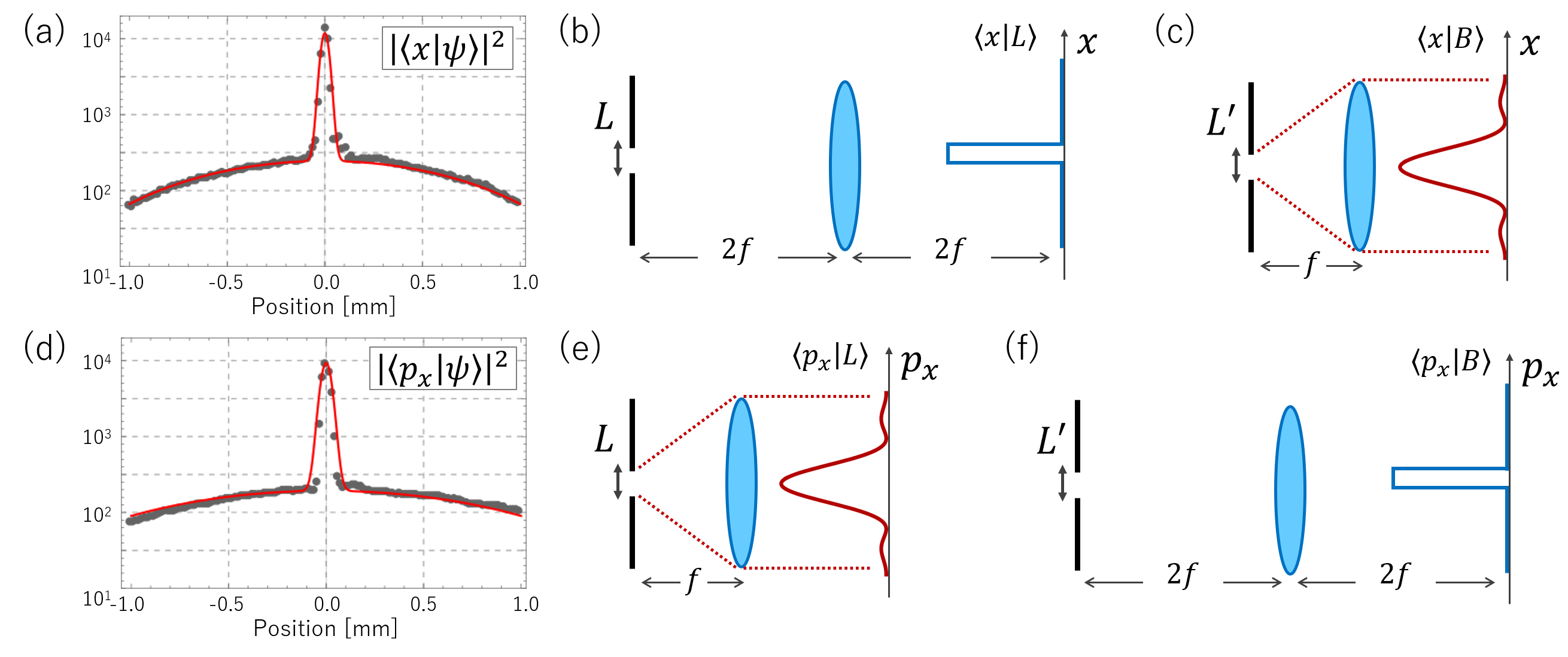}
 \caption{(a) Interference fringe of position and momentum obtained experimentally in position space for initial state of $|\psi \rangle$. Red is the fitting curve. The fitting was done using the Gaussian function instead of the slit function. (b) is the measurement system for the position state in position space, (c) is the measurement system for the momentum state in position space.
(d) Interference fringe of position and momentum obtained experimentally in momentum space for initial state of $|\psi  \rangle$. (e) is the measurement system for the position state in the momentum space, and (f) is the measurement system for the momentum state in the momentum space.
}
 \label{fig4}
\end{figure*}

Figure 3 (b) shows the interference fringe at intermediate time $z=1.4 \times z_M$ which is slightly longer than its optimal time of $1.1 \times z_M$. 
We have clearly observed an interference fringe between position and momentum state. 
From the fitting, measured visibility was roughly 85\%.
Figure 4(a) and 4(d) show the measured probability densities at $z=0$ in position basis and momentum basis respectively.
For the measurement in position basis, the interference fringe was taken on the image plane of the position slit $L$ outside of the interferometer as shown in fig.4(b) and (c).
The probability of $P(L)$ is then calculated by integrating the probability density at $z=0$ over the position $x$ from $-L/2$ to $L/2$.
From the fitted curve, the probability was calculated as $P(L) = 56.5 \%$.
For the measurement in momentum basis, we used the effect on Fourier transform by using the lens, where the state of eq. (\ref{superposition}) is mapped into momentum space of $\hat{p}_x$.
For position state of $\langle p_x | L \rangle$, we put the lens at the position after the slit at the distance of focal length $f$ of the lens, where the momentum information before the lens is mapped onto the position scaled by $h/(f\lambda)$ after the lens.
For momentum state of $\langle p_x | B \rangle$, we put the lens at the position after the slit at the distance of $2f$.
The interference fringe was then taken on the image plane of the position slit $L'$ outside of the interferometer as shown in fig.4(e) and (f).
The probability of $P(B)$ was then calculated by integrating the probability density at $t=0$ over the momentum from $-B/2$ to $B/2$.
From the fitted curve, the probability was calculated as $P(B) = 56.5 \%$.

Finally, we analyse the data.
At initial position of $t=0$, we obtained the probabilities of $P(L)$ and $P(B)$ as 56.5 \%.
We also measured the probability at intermediate position.
From the propagation inequality of eq.(\ref{propagation_ineq}), the minimum value of the probability of finding the photon within a interval $M$ at $t=1.4 \times t_M$ was predicted as $P(L) + P(B) - 1 \approx 13.1$ \% which corresponds to the area under the purple line as shown in fig.3(b).
The figure clearly shows that the actual measured probability of $P(M, t)$ is smaller than the value.
The value of $P(M = L + 1.4 \times Bt_M/m)$ is calculated by integrating the probability density at $t=1.4 \times t_M$ over the position $x$ from $-M/2$ to $M/2$.
From the fitting, which takes into account the visibility of the fringe, the probability was calculated as $P(M, t=1.4\times t_M) = 7.2 \%$.
From the measured data, we obtained $P_{\mathrm{defect}}(t=1.4 \times t_M) \approx 5.9 \%$ which means that 5.9 \% probability cannot be explained by the statistical limit which assume that particle move along straight line.

In conclusion, we have experimentally generated superposition states of position and momentum using photons, interferometers, slits and lenses. The strong interference of position and momentum was obtained by precisely adjusting the slit width, lens and the propagation distance of the photons. 
Using the generated state, we predicted a lower limit on the probability of finding a particle at an intermediate position, based on the statistical information on the position and momentum of the particle at initial time.
We then measured the probability of finding the particle at the intermediate position.
The results show that 5.9 \% probability cannot be explained by the statistical limit which assume that particle move along straight line.
In this proof-of-principle experiment, we used weak coherent (laser) light attenuated to the single photon level. A single-photon source would further improve the clarity, since there would be no multiphoton contribution.

\nocite{*}
This work was supported by the EPSRC Quantum Technology Hub in quantum imaging QUANTIC and the Centre for Nanoscience and Quantum Information (NSQI). T.O. is supported by JST, PRESTO.

\bibliography{references}

\end{document}